\documentclass[aps,prl,reprint,superscriptaddress]{revtex4-2}
\usepackage{amsmath,amsfonts,amssymb,bm,hyperref,graphicx,subfigure,orcidlink}
\usepackage{changes}
\usepackage[notrig]{physics}
\usepackage{CJKutf8}
\usepackage{xcolor}
\usepackage{todonotes}
\UseRawInputEncoding
\newcommand{\ud}{\mathrm{d}}

\newcommand{\ue}{\mathrm{e}}
\newcommand{\ui}{\mathrm{i}}

\begin{document}

\begin{CJK*}{UTF8}{gbsn} 
\title{Extending Covariant Fluctuation Theorems into Quantum Regime\\ through Quasiprobability Approach}

\author{Ji-Hui Pei (裴继辉)\orcidlink{0000-0002-3466-4791}}
\thanks{These authors contribute equally.}
\affiliation{School of Physics, Peking University, Beijing, 100871, China}
\affiliation{Department of Physics and Astronomy, KU Leuven, 3000, Belgium}

\author{Tingzhang Shi (石霆章)\orcidlink{0009-0005-4283-9669}}
\thanks{These authors contribute equally.}
\affiliation{School of Physics, Peking University, Beijing, 100871, China}

\author{Jin-Fu Chen (陈劲夫)\orcidlink{0000-0002-7207-969X}}
\email{jinfuchen@lorentz.leidenuniv.nl}
\affiliation{Instituut-Lorentz, Universiteit Leiden, P.O. Box 9506, 2300 RA Leiden, The Netherlands}
\affiliation{$\langle aQa^L\rangle$ Applied Quantum Algorithms Leiden, The Netherlands}
\author{H. T. Quan (全海涛)\orcidlink{0000-0002-4130-2924}}
\email{htquan@pku.edu.cn}
\affiliation{School of Physics, Peking University, Beijing, 100871, China}
\affiliation{Collaborative Innovation Center of Quantum Matter, Beijing 100871, China}
\affiliation{Frontiers Science Center for Nano-optoelectronics, Peking University,
Beijing, 100871, China}

\begin{abstract}
The covariant formulation of stochastic thermodynamics requires treating the stochastic work as a 4-vector, posing significant challenges for quantum systems due to the non-commutativity. 
We introduce a new quasiprobability distribution for the work 4-vector, which combines the Wigner and Margenau-Hill quasiprobabilities. 
This extends the covariant fluctuation theorems from classical to quantum regime.
We illustrate our findings with a scalar field driven by classical particles with a generalized version of trace formula. 
Our work establishes a quasiprobability approach to studying relativistic
quantum thermodynamics 
in a covariant way.
\end{abstract}

\maketitle
\end{CJK*}

\textit{Introduction.}
Fluctuation theorems \cite{Jarzynski1997,Crooks1998,Jarzynski2011} can be viewed as a generalization of the second law of thermodynamics for small systems, which bridge irreversibility and fluctuations through thermodynamic quantities. 
Although they were firstly derived for closed or open systems that obey Newtonian or Langevin dynamics, 
they can be
generalized to quantum systems \cite{Tasaki2000,Kurchan2001,Talkner2007,Esposito2009,Campisi2011}, special relativistic systems \cite{Fingerle2007,Cleuren2008,Pal2020,Cai2023a,Cai2023,Zhang2024},
as well as relativistic quantum fields \cite{Fei2019,Ortega2019,Bartolotta2018,TeixidoBonfill2020,Torrieri2021}.

The principle of covariance is a cornerstone of modern physics and is a central topic in special relativity. 
However, most previous studies of fluctuation theorems are restricted to the rest reference frame and do not incorporate the principle of covariance. 
Covariant fluctuation theorems for classical systems \cite{Pei2025,Shi2025} are recently proposed by adopting the concept of work 4-vector and heat 4-vector which incorporate both energy and momentum components. 
These covariant fluctuation theorems are valid for any inertial observer and can be applied to arbitrary classical thermodynamic systems. 

A natural idea is to extend the covariant fluctuation theorems into the quantum regime. 
However, this extension faces a fundamental obstacle: in relativistic quantum systems, the Hamiltonian and momentum operators generally do not commute. Therefore, the components of the 4-momentum cannot, in general, be measured simultaneously, and a joint probability distribution for the work 4-vector cannot be straightforwardly defined.
For quantum systems, how to incorporate fluctuation theorems with the principle of covariance is one of the most challenging problems.

Quasiprobability distributions, which may take negative or complex values, arise naturally when treating noncommutative observables in quantum physics.
Known examples include Wigner distribution \cite{Wigner1932,Luks1997}, Kirkwood-Dirac (KD) distribution \cite{Kirkwood1933,Dirac1945,ArvidssonShukur2024}, and the real part of the latter, Margenau-Hill (MH) distribution \cite{Margenau1961}. 
Quasiprobability distributions have recently found many applications in quantum thermodynamics \cite{Levy2020,Lostaglio2023,HernandezGomez2024,Gherardini2024,Mallik2025,SilvaPratapsi2025,Yoshimura2026,Jae2026}. 
For example, for a driving process in which the initial state does not commute with the Hamiltonian,
it is impossible to define a satisfactory probability distribution for quantum work \cite{PerarnauLlobet2017,Hovhannisyan2024}.
Nevertheless, the quantum work defined through quasiprobability distribution provides a solution to this problem \cite{Lostaglio2018,Francica2022,Francica2022a,Lostaglio2023,Pei2023}.

In this Letter, for driven relativistic quantum fields, 
we introduce a new quasiprobability distribution of work 4-vector which combines the Wigner and MH ones. 
This distribution is proved to be unique under two requirements: preserving the principle of covariance and recovering the MH distribution for the energy component. A quantum version of the covariant Jarzynski equality is found to be valid for such a distribution. However,
different from the classical case \cite{Pei2025}, the quantum version of the covariant Crooks fluctuation theorem is found to be valid only for a specific linear combination and is violated for the joint quasiprobability distribution.
As an example, we explicitly study the statistics of work 2-vector for a (1+1)D scalar quantum field driven by classical particles.
We thus provide a framework for studying
fluctuations and irreversibility of quantum systems in a covariant way.

\textit{Setup.}
We consider the following finite-time thermodynamic process: 
the system is an isolated relativistic quantum field $\phi(x)$, which is driven by an external driving field $h(x)$. 
We assume the driving process starts at an initial spacelike hypersurface $\Sigma_-$ and ends at a final spacelike hypersurface $\Sigma_+$. 
The driving field, as a pre-determined classical field, only changes in between the two hypersurfaces. 
In the following, we use Einstein's summation rule with the metric $\eta_{\mu\nu}=\operatorname{diag}(1,-1,-1,-1)$ 
and adopt the Heisenberg picture.

To build a Lorentz covariant theory, we should use quantities with well-defined transformation rules between different inertial reference frames.
We start with the energy-momentum tensor $T^{\mu\nu}$ of the system, which should be treated as a quantum operator. 
Since the field is subject to an external driving field, $T^{\mu\nu}$ is no longer a conserved current.
The flux of the energy-momentum tensor on the initial and final hypersurfaces yields the initial and final 4-momentum operators,
\begin{equation}\label{eq:four-momentum}
    P_\mp^\nu=\int_{\Sigma_\mp}\ud\sigma n_\mu T^{\mu\nu}.
\end{equation}
The above formula represents a surface integral on $\Sigma_\mp$, where $\ud \sigma$ is the surface element, and $n_\mu$ denotes the normal vector. 

We suppose that the system is initially in equilibrium and is later driven out of equilibrium. 
We denote the rest inverse temperature of the initial state of the system as $\beta$. 
In an arbitrary inertial reference frame, one observes a moving equilibrium state at 4-velocity $u^\mu$. 
The density matrix can be formally written in the covariant form \cite{Hakim2011},
\begin{equation}\label{rhoeq}
    \rho^\mathrm{eq} = \exp(-\beta_\mu P^\mu_-) / Z_-,
\end{equation}
with the inverse temperature 4-vector $\beta_\mu = \beta u_\mu$. 
The partition function $Z_-=\Tr  \exp(-\beta_\mu P^\mu_-)$ is related to the free energy $F_-$ by $Z_- = \exp(-\beta F_-)$. 
In the rest frame $u_\mu= (1,0,0,0)$, the initial state reduces to the conventional form $\rho^\mathrm{eq}=\exp(-\beta H_-)/Z_-$. 
For a field system, the precise definition of a moving equilibrium state is via the Kubo-Martin-Schwinger condition \cite{Kubo1957,Martin1959}. 
Nevertheless, in our framework, it is safe to use the formal density matrix in Eq.~\eqref{rhoeq}.

To explore the irreversibility requires the concept of the backward process, in which quantities are denoted by an overbar. 
In the covariant formalism, the backward process is defined as the spacetime reversal of the forward process instead of only the time reversal. 
The initial and final hypersurfaces in the backward process are respectively spacetime reversals of the final and initial hypersurfaces 
in the forward process, i.e., $\bar\Sigma_- = -\Sigma_+$, $\bar \Sigma_+=-\Sigma_-$.
Meanwhile, the driving field during the process undergoes a spacetime reversal, 
which, depending on the parity under spacetime reversal, is given by $\bar h(x)=\pm h(-x) $. 
Note that the 4-velocity  does not change under spacetime reversal. 
For the initial state of the backward process, 
we consider an equilibrium state at the same inverse temperature 4-vector $\beta_\mu$,
\begin{equation}
    \bar \rho^\mathrm{eq} = \exp(-\beta_\mu\bar P^\mu_-) / Z_+,
\end{equation}
where $Z_+=\Tr \exp(-\beta_\mu\bar P^\mu_-)$ is the partition function of the initial state of the backward process.
The initial 4-momentum $\bar P^\mu_-$ in the backward process is generally different from that in the forward process, 
due to the different initial configuration of the driving field $\bar h(x)\eval_{\bar\Sigma_-}$.

For a classical relativistic driving process, 
it is possible to introduce a work 4-vector $W^\mu$ associated with a stochastic trajectory, 
which, for an isolated driven system, is equal to the 4-momentum change between the final and initial states. 
Then, the joint distribution of the work 4-vector $P_c(\{w^\mu\})$ in the forward and backward processes satisfies the following covariant Crooks fluctuation theorem \cite{Pei2025}
\begin{equation}\label{classical}
    \frac{P_\text{c}(\{w^\mu\})}{\bar P_\text{c}(\{-w^\mu\})} = \exp(\beta_\mu w^\mu -\beta\Delta F),
\end{equation}
where $\Delta F =-\ln(Z_{+}/Z_{-})/\beta$ is the free energy difference between two initial equilibrium states.

However, this formula cannot be directly applied to quantum systems 
because the work 4-vector associated with a trajectory is not well-defined in quantum regime. 
Even worse, energy and momentum operators generally do not commute with each other,
leading to the failure of standard definitions in quantum thermodynamics, 
such as the two-projective measurement scheme and the Ramsey scheme \cite{Dorner2013,Mazzola2013,Batalhao2014,Ortega2019}.

\textit{Quasiprobability of quantum work.}
As is known, two noncommutative operators $A$ and $B$ cannot be measured at the same time, and one cannot define a joint probability distribution.
The quasiprobability approach provides a joint distribution $P(a,b)$, at the expense of negative-valued or complex-valued distribution. 
In the End Matter, , we review several different quasiprobability distributions, including the Wigner distribution, the KD distribution, and the MH distribution.

The difficulty in defining the work 4-vector due to noncommutativity is twofold. 
First, for an initially moving equilibrium state, the density matrix does not commute with either the momentum or the energy operators,
making the measurement invasive. 
Second, in general, the energy and momentum operators do not commute, preventing simultaneous counting. 

The first one traces back to a famous conceptual problem in the (non-covariant) quantum thermodynamics: 
how to define the quantum work distribution for a general initial state, especially when the initial density matrix does not commute with the energy. 
In this context, work corresponds to the 0th component of the work 4-vector in the covariant formulation, and the standard definition through the two-projective-measurement scheme is invalid as the first energy measurement destroys the initial state. 
Furthermore, a decisive no-go theorem \cite{PerarnauLlobet2017,Hovhannisyan2024} states that there is no way to define a satisfactory distribution for quantum work within the framework of standard probability distribution.

Fortunately, this conundrum can be solved by resorting to quasiprobability.  
It has been shown that, the MH quasiprobability of work \cite{Allahverdyan2014} is the only advantageous definition since it satisfies all the physically reasonable requirements \cite{Pei2023,Yi2025}. 
In this approach, one first defines the joint MH distribution of the initial and final Hamiltonian, $H_-$ and $H_+$. 
They are in the Heisenberg picture, so that time evolution has been taken into account.
The work distribution is identified as the distribution of $H_+-H_-$ induced by their joint quasiprobability distribution. 
This MH quasiprobability of work is real-valued but not positive definite. 
Through this definition, work is associated with the energy change between the initial and final states.

Now that the first difficulty has a clear solution, the main problem we try to solve is the second one: 
the energy and momentum operators do not commute and thus cannot be counted simultaneously.
This urges us to introduce a new quasiprobability distribution.

\textit{Quasiprobability distribution of work 4-vector.}
We seek a quasiprobability distribution that satisfies the following requirements: 
(1) Under Lorentz transformation $x^{\prime\mu} =\Lambda^\mu_\nu x^\nu$, the probability distribution functions in two inertial frames satisfy $P_\text{WMH}(\{w^\mu\})=P^\prime_\text{WMH}(\{\Lambda^\mu_\nu w^\nu\})$. 
(2) In an arbitrary inertial reference frame, the marginal distribution of the energy component $W^0$ reduces to the MH quasiprobability distribution of work. 
The first requirement is from the principle of covariance, and the second one is 
based on the previous knowledge about the non-covariant quantum work for a generic initial state. 

The distribution satisfying both requirements is almost unique, 
given by a mixed distribution of Wigner and MH (WMH) quasiprobability \cite{[{See }] [{ for detailed theoretical derivations.}] SM}.
The corresponding joint characteristic function of work 4-vector is defined as 
\begin{equation}\label{WMH}
\begin{aligned}
    \chi_\text{WMH}(\{\tilde w_\mu\}) =& \frac12\Tr\left[\exp(\ui\tilde w_\mu P^\mu_+)\exp(-\ui\tilde w_\mu P^\mu_-) \rho\right]\\
    +&\frac12\Tr\left[\exp(-\ui\tilde w_\mu P^\mu_-)\exp(\ui\tilde w_\mu P^\mu_+) \rho\right],
    \end{aligned}
\end{equation}
where $P^\mu_-$ and $P^\mu_+$ are the initial and final 4-momentum. 
The joint quasiprobability distribution $P_\text{WMH}(\{w^\mu\})$ is obtained from the inverse Fourier transform of $\chi_\text{WMH}(\{\tilde w_\mu\})$. 
The definition in Eq.~\eqref{WMH} distinguishes operators at the same time (or hypersurface) and operators at different times (or hypersurfaces). 
For operators at the same time, they are counted in the Wigner way, i.e., summed in the exponential, while for operators at different times, they are counted in the MH way, i.e., time ordered.

This WMH quasiprobability $P_\text{WMH}(\{w^\mu\})$ is real but not positive-definite, and the distribution of any linear combination of $W^\mu$ reduces to the MH distribution.  
It has desirable properties in its first and second order moments: 
(1) The average value of the work 4-vector is equal to the average value of the 4-momentum change.
(2) The covariance matrix is positive.

\textit{Covariant quantum fluctuation theorems.}
We now investigate whether this quasiprobability distribution of work 4-vector satisfies fluctuation theorems. 
Recall our setup: a quantum field is initially in a moving equilibrium state at the inverse temperature 4-vector $\beta_\mu$, given by Eq.~\eqref{rhoeq}, and the system is then driven out of equilibrium by an external field. 
We denote the WMH quasiprobability distributions in the forward and backward processes as $P_\text{WMH}(\{w^\mu\})$ and $\bar P_\text{WMH}(\{w^\mu\})$, respectively.

By choosing $\tilde w_\mu =i\beta_\mu$ and $\rho=\rho^\text{eq}$ in Eq.~\eqref{WMH}, we obtain $\chi_{\mathrm{WMH}}(\{i\beta_\mu\})=Z_+/Z_-$, and therefore the WMH quasiprobability fulfills the covariant Jarzynski equality, 
\begin{equation}
   \langle e^{-\beta_{\mu}W^{\mu}}\rangle =\int\ud^{4}wP_{\text{WMH}}(\{w^{\mu}\})e^{-\beta_{\mu}w^{\mu}}=e^{-\beta\Delta F}.
\end{equation}
By using Jensen's inequality, we obtain the maximum work principle at the ensemble average level, 
\begin{equation}
    \ev{\beta_\mu W^\mu} -\beta\Delta F \geq 0.
\end{equation}
The above results are valid for any inertial reference frame. 
It indicates that the irreversibility in a driving process is 
characterized by a Lorentz scalar $\beta_\mu W^\mu -\beta\Delta F$.

One may expect a quantum counterpart of Eq.~\eqref{classical}, 
but unfortunately, the covariant version of Crooks fluctuation theorem does not hold true at the level of joint distribution unless the energy operator and the momentum operators are commutative.
That is, we have in general
\begin{equation}
    \frac{P_\text{WMH}(\{w^\mu\})}{\bar P_\text{WMH}(\{-w^\mu\})} \neq \exp\left( \beta_\mu w^\mu -\beta\Delta F\right).\label{eq:violationcrook}
\end{equation}
Nevertheless, 
we may recover the Crooks fluctuation theorem at the level of the linear combination $u_\mu W^\mu$, whose one-dimensional distribution $P_{u_\mu W^\mu}(w)$ is induced by the WMH distribution,
and it satisfies 
\begin{equation}
    \frac{P_{u_\mu W^\mu}(w)}{\bar P_{u_\mu W^\mu}(-w)} = \exp\left( \beta w -\beta\Delta F\right).\label{eq:crooksfluctuationtheoremlinearcombination}
\end{equation}
Note that $u_\mu W^\mu = W^0_\text{rest}$ is 
the energy component of the work 4-vector in the rest reference frame,
and $P_{u_\mu W^\mu}(w)$ reduces to the two-projective-measurement definition of the positive definite work distribution \cite{Tasaki2000, Kurchan2001,Talkner2007} in the rest frame. 
 
The breakdown of Crooks fluctuation theorem at the level of joint distribution 
marks the key difference between the classical and quantum thermodynamics. 
Since the WMH quasiprobability is unique under the physically reasonable requirements \cite{Pei2023,SM}, one cannot get around such a problem by choosing another quasiprobability distribution of work 4-vector. 
For noncommutative variables, 
one can only obtain a detailed fluctuation theorem at the level of distribution of a linear combination $u_\mu W^\mu$.

\textit{Open systems.}
For an open quantum system with relative motion to the heat bath, we cannot find a rest reference frame for both the system and the heat bath, and 
it is thus desirable to build a covariant formalism of heat and work 4-vectors.

By considering the case where some of the quantum fields are treated as the system while the others are treated as the heat bath, the formalism of covariant fluctuation theorems can be generalized to open quantum systems. 
Treating the system fields plus the bath fields as a composite isolated quantum system and assuming the interaction between the system and the bath is weak,
we can follow the standard procedure for thermodynamics of open quantum systems \cite{Talkner2009a,Esposito2009,Campisi2011}.
In the weak-interaction regime, the 4-momentum of the system and the 4-momentum of the bath commute with each other. 
Work and heat 4-vectors are identified as the 4-momentum change of the composite system and the heat bath, respectively. 
We may define a joint distribution of work and heat 4-vectors by using the mixed Wigner and MH quasiprobability, whose characteristic function reads
\begin{equation}
\begin{aligned}
    &\chi_\mathrm{WMH}(\{\tilde w_\mu\},\{\tilde q_\mu\}) =\\
    &\frac12\Tr[\ue^{\ui \tilde w_\mu P^\mu_{\text{t}+}-\ui \tilde q_\mu P^\mu_\text{b+}}\ue^{-\ui \tilde w_\mu P^\mu_{\text{t}-}+\ui \tilde q_\mu P^\mu_{\text{b}-}}\rho]\\
    +&    \frac12\Tr[\ue^{-\ui \tilde w_\mu P^\mu_{\text{t}-}+\ui \tilde q_\mu P^\mu_{\text{b}-}}\ue^{\ui \tilde w_\mu P^\mu_{\text{t}+}-\ui \tilde q_\mu P^\mu_\text{b+}}\rho],
    \end{aligned}
\end{equation}
where $P^\mu_{\text{t}+}$ and $P^\mu_\text{b+}$ are the final 4-momentum of the composite system and the heat bath, while $P^\mu_{\text{t}-}$ and $P^\mu_{\text{b}-}$ are the initial 4-momentum of the composite system and the heat bath. 

In a driving process where the initial temperatures of the system and the heat bath are $\beta^\text{s}_\mu=\beta^\mathrm{s}u_\mu^\text{s}$ and $\beta^\text{b}_\mu=\beta^\text{b}u_\mu^\text{b}$, we obtain a generalized covariant Crooks fluctuation theorem \eqref{eq:crooksfluctuationtheoremlinearcombination} in terms of two linear combinations, the system energy change measured in the rest frame of the system $u_\mu^\text{s} (W^\mu +Q^\mu)$ and the bath energy change measured in the rest frame of the bath $u_\mu^\text{b}Q^\mu$.
The covariant Crooks fluctuation theorem is
\begin{equation}
\begin{aligned}
    &\frac{P_{u_\mu^\text{s} (W^\mu +Q^\mu),u_\mu^\text{b}Q^\mu}(u_\mu^\text{s}(w^\mu+q^\mu),u_\mu^\text{b}q^\mu)}{\bar P_{u_\mu^\text{s} (W^\mu +Q^\mu),u_\mu^\text{b}Q^\mu}(-u_\mu^\text{s}(w^\mu+q^\mu),-u_\mu^\text{b}q^\mu)} \\
    = &\exp\left[\beta_\mu^\text{s} w^\mu-\beta^\mathrm{s}\Delta F-(\beta^\text{b}_\mu-\beta^\text{s}_\mu)q^\mu\right] ,
    \end{aligned}\label{eq:covariantcrooksfluctuation}
\end{equation}
where $\Delta F$ is the free energy difference of the system. We show the derivation to Eq.~\eqref{eq:covariantcrooksfluctuation} in~\cite{SM}. When both the system and the heat bath are at rest, Eq.~\eqref{eq:covariantcrooksfluctuation} recovers the fluctuation theorems for the joint distribution of work and heat \cite{Talkner2009a,Chen2023}.

\textit{Scalar field driven by classical particles.}
\begin{figure*}
    \centering
    \includegraphics[width=\linewidth]{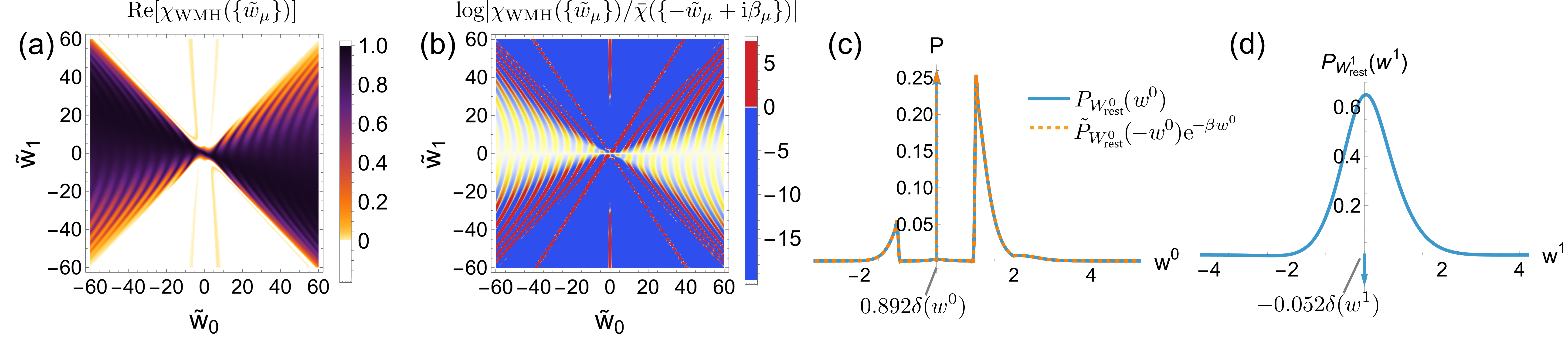}
    \caption{Numerical results of the characteristic function and distribution function. The parameters in the protocol are chosen as $\beta=1.5,v_b=0.5,T=1.0,\sigma=1.0,gN_0=3.5$.
    (a) The real part of the characteristic function of the joint WMH quasiprobability distribution of the work 2-vector. (b) The logarithmic ratio between $\chi_{\text{WMH}}(\{\tilde w_\mu\})$ and $\bar \chi_\text{WMH} (\{-\tilde w_\mu+\ui \beta_\mu\})$. The deviation from $0$ indicates the breakdown of the Crooks fluctuation theorem (Eq.~\eqref{eq:violationcrook}). (c) The marginal distribution of the energy component $P_{W_{\mathrm{rest}}^0}(w^0)$ in the forward process and the corresponding $\bar{P}_{W_{\mathrm{rest}}^0}(-w^0)\exp(-\beta w^0)$ in the reverse process. The probability density consists of a continuous distribution and a delta distribution at $w^0=0$. (d) The marginal distribution of momentum component $P_{W_{\mathrm{rest}}^1}(w^1)$. It consists of a continuous distribution and a delta distribution at $w^1=0$. The magnitude of the delta distribution is negative, indicating the quasiprobability feature.}
    \label{fig: all}
\end{figure*}
To illustrate our result, we explicitly calculate the joint quasiprobability distribution of the work 2-vector in a simple model: A (1+1)D scalar free field 
$\phi$ driven by a beam of classical particles. 
The Lagrangian of the system reads
\begin{equation}
    \mathcal{L}=\frac{1}{2} \partial^\mu\phi\partial_\mu\phi-\frac{1}{2}m^2 \phi^2-h\phi,
\end{equation}
with the external potential generated by the classical particles,
\begin{equation}
    h(x^\mu)=g\int \ud l \int \ud\tau~n(l) \delta^2[x^\mu-x^\mu_l(\tau)].
\end{equation}
Here $g$ is the coupling strength, $n(l)$ represents the particle number density over the continuous variable $l$, which labels different particles, 
and $x^\mu_l(\tau)$ is the predetermined worldline with proper time $\tau$ of the $l$-labeled particle. 

We consider the following linear quench process in the rest frame of the system.
The initial and final hypersurfaces $\Sigma_\mp$ are chosen to be the isochronous planes at $t=0$ and $t=T$.
The particles are initially at rest relative to the system, satisfying a Gaussian density distribution $n(l)=N_0/(\sqrt{2\pi}\sigma)\exp(-l^2/2\sigma^2)$ over the initial spatial coordinate $l$, where $N_0$ is the total number of the particles, and $\sigma$ is the spread of the beam.
At initial time $t=0$, every particle in the beam instantaneously acquires a velocity $v_b$ and subsequently moves with this velocity.
At final time $t=T$, particles suddenly become static again.
Accordingly, the trajectory of each particle can be written as 
\begin{equation}
x_l^1(t)=
\begin{cases}
l , &\quad t\leq0\\
l+v_b t , &\quad 0<t<T\\
l + v_b T ,&\quad t\geq T.
\end{cases}
\end{equation}
During this process, every particle changes its coordinate by $v_b T$.

Since the translational symmetry is broken, energy and momentum do not commute with each other.
In the Heisenberg picture, the initial and final 4-momentum operators $P^\mu_{\mp}$ of mode $k$ can be solved
in terms of annihilation and creation operators $(a_k,a^\dagger_k)$,
\begin{equation}
    P^\mu_{\mp}=\int \frac{\ud k}{4\pi\omega_k} [k^\mu a^\dagger_k a_k+A^\mu_{\mp}(k)a_k+A^{\mu *}_{\mp}(k)a^\dagger_k+B^\mu_{\mp}(k)],
\end{equation}
which are quadratic forms with linear and constant terms \cite{SM}. 
$A^\mu_{\mp}(k)$ and $B^\mu_{\mp}(k) $ are coefficients, and $k^\mu =(\omega_k =\sqrt{k^2+m^2},k)$.
We can then calculate the characteristic function of the work 2-vector with Eq.~(\ref{WMH}) using the trace formula~\eqref{eq:traceformula} in the End Matter.

In the rest frame, the numerical result of the characteristic function is shown in Fig.~\ref{fig: all}(a).
Since the characteristic function is a Lorentz scalar, in a generic inertial reference frame, the plot of the characteristic function will undergo a Lorentz boost in its input $\tilde w_\mu$.
The deviation from the Crooks fluctuation theorem for the WMH characteristic function is shown in Fig.~\ref{fig: all}(b).
From the numerical results of the characteristic function, we obtain the marginal distribution of $W^0_{\mathrm{rest}}$ and $W^1_{\mathrm{rest}}$.
By comparing $P_{W^0_{\mathrm{rest}}}(w^0) \ue^{-\beta w^0}$ and $\bar{P}_{W^0_{\mathrm{rest}}}(-w^0)$, we verify the Crooks fluctuation theorem for $W^0_{\mathrm{rest}}$; see Fig.~\ref{fig: all}(c).
We also find that the marginal distribution of $W^1_\text{rest}$ can be decomposed into a continuous distribution and a delta distribution at $w^1=0$ with a negative magnitude, illustrating the feature of quasiprobability distribution; see Fig.~\ref{fig: all} (d).

\textit{Conclusions.}
The covariant formalism of quantum thermodynamics requires one to incorporate the energy and momentum change into a work 4-vector. 
We introduce a new quasiprobability, which mixes the Wigner and MH ones, to define the joint distribution of work 4-vector in quantum field systems. 
This quasiprobability distribution has a series of desirable properties in its first and second order moments, 
including the first law of thermodynamics at the average level and the positivity of the covariance.
Although the covariant Jarzynski equality is satisfied, the covariant Crooks fluctuation theorem is valid only for the distribution of a specific linear combination of the work 4-vector. 
This indicates a clear difference between the quantum and classical covariant fluctuation theorems.
Our results not only build a covariant theory of nonequilibrium thermodynamics for quantum relativistic systems,
but also establish a new quasiprobability framework to characterize the joint distributions of noncommutative variables.

\begin{acknowledgments}
\textit{Acknowledgments.}
This work is supported by the National Natural
Science Foundation of China (NSFC) under Grants No. 12375028, No. 12521004. J.F.C. also
acknowledges the support received from the
European Union’s Horizon Europe research and
innovation programme through the ERC StG
FINE-TEA-SQUAD (Grant No. 101040729).
The views and opinions expressed here are
solely those of the authors and do not necessarily
reflect those of the funding institutions. None of
the funding institutions can be held responsible
for them.
\end{acknowledgments}

\bibliography{cqft}

\clearpage
\section{End Matter}

\textit{Quasiprobability approach.}
For each quasiprobability distribution, there is a one-to-one correspondence between the characteristic function $\chi(\tilde a,\tilde b)$ and the distribution function via the Fourier transform,
\begin{equation}
    \chi(\tilde a,\tilde b) = \int\ud a\ud b \exp(\ui a \tilde a+\ui b\tilde b) P(a,b).
\end{equation}

In the following, we review several different quasiprobability distributions, including the Wigner distribution $P_\text{Wi}(a,b)$, the KD distribution $P_\text{KD}(a,b)$, and the MH distribution $P_\text{MH}(a,b)$. 
Their characteristic functions are given by 
\begin{align}
    &\chi_\text{Wi}(\tilde a,\tilde b) = \Tr\left[\ue^{\ui\tilde aA+\ui\tilde bB} \rho\right],\\
    &\chi_\text{KD}(\tilde a,\tilde b) = \Tr\left[\ue^{\ui\tilde aA} \ue^{\ui\tilde bB} \rho\right],\\
    &\chi_\text{MH}(\tilde a,\tilde b) 
            =\frac12\Tr\left[(\ue^{\ui\tilde aA} \ue^{\ui\tilde bB}+\ue^{\ui\tilde bB}\ue^{\ui\tilde aA}) \rho\right].
\end{align}
In the characteristic function of the Wigner quasiprobability, two operators $A$ and $B$ are summed together in the exponential, 
and the corresponding distribution function $P_\text{Wi}(a,b)$ is real-valued but can be negative. 
For an arbitrary linear combination of $A$ and $B$, 
the Wigner quasiprobability provides the correct marginal probability distribution. 
In the characteristic function of the KD quasiprobability, two operators $A,B$ are ordered separately, and the ordering of them matters. In contrast, the MH quasiprobability involves a symmetric ordering of $A$ and $B$.
The KD distribution $P_\text{KD}(a,b)$ is generally complex-valued, and the MH distribution is its real part, $P_\text{MH}(a,b) = \Re P_\text{KD} (a,b)$. 
For the KD and MH quasiprobabilities, 
their distribution functions are nonzero only if both $a$ and $b$ are the eigenvalues of $A$ and $B$. 
All the above quasiprobability distributions reduce to the ordinary probability distribution when $A$ and $B$ commute.

\textit{Trace formula.}
Here, we give the trace formula for the products of exponential Bosonic quadratic forms (with linear terms), 
which is used in the calculation of the characteristic function of work by Eq.~\eqref{WMH} in our example. 

We consider a Bosonic system with $N$ modes and use the Nambu basis,
\begin{equation}
    A = (a_1,\cdots,a_N,a_1^\dagger,\cdots,a_N^\dagger).
\end{equation}
The commutation relation is
\begin{equation}
    \comm{A_i}{A_j}= \Omega_{ij} = 
    \begin{pmatrix}
        0 & I\\
        -I & 0
    \end{pmatrix}_{ij}.
\end{equation}
Consider several operators up to quadratic terms, 
\begin{equation}
    H_i = \frac12A^T G_i A+l_i^TA + h_i,
\end{equation}
with matrices $G_i$ as coefficients for quadratic terms, vectors $l_i^T$ as coefficients for linear terms, and constants $h_i$. 
The trace formula of the product of their exponentials is given by
\begin{equation}
\begin{aligned}
    &\Tr\left[\ue^{-H_{1}}\cdots\ue^{-H_{n}}\right]\\=&\exp(\sum_{i}\frac{1}{2}l_{i}^{T}\frac{S_{i}-\sinh S_{i}}{S_{i}^{2}}\Omega l_{i}-h_{i})\exp(\frac{1}{2}\sum_{i<j}d_{i}^{T}\Omega d_{j})\\&\times\exp[\frac{1}{4}d^{T}\Omega(\coth\frac{S}{2})d][(-)^{N}\det(I-M)]^{-\frac{1}{2}},
    \end{aligned}\label{eq:traceformula}
\end{equation}
where
\begin{equation}
\begin{aligned}
&S_{i}=\Omega G_{i},\\
 &   d_{i}=-\ue^{-S_{i}}\cdots\ue^{-S_{i-1}}(I-\ue^{-S_{i}})S_{i}^{-1}\Omega l_{i},\\
  &  S = -\log \ue^{-S_1}\cdots\ue^{-S_n},\quad d=d_1+\cdots d_n.
    \end{aligned}
\end{equation}
We remark that Eq.~\eqref{eq:traceformula} is a generalization of the results in Ref.~\cite{Fei2019a}.
\end{document}


\begin{CJK*}{UTF8}{gbsn} 

\title{Supplemental Material:\\ Extending Covariant Fluctuation Theorems into Quantum Regime through Quasiprobability Approach}

\author{Ji-Hui Pei (裴继辉)\orcidlink{0000-0002-3466-4791}}
\thanks{These authors contribute equally.}
\affiliation{School of Physics, Peking University, Beijing, 100871, China}
\affiliation{Department of Physics and Astronomy, KU Leuven, 3000, Belgium}

\author{Tingzhang Shi (石霆章)\orcidlink{0009-0005-4283-9669}}
\thanks{These authors contribute equally.}
\affiliation{School of Physics, Peking University, Beijing, 100871, China}

\author{Jin-Fu Chen (陈劲夫)\orcidlink{0000-0002-7207-969X}}
\email{jinfuchen@lorentz.leidenuniv.nl}
\affiliation{Instituut-Lorentz, Universiteit Leiden, P.O. Box 9506, 2300 RA Leiden, The Netherlands}
\affiliation{$\langle aQa^L\rangle$ Applied Quantum Algorithms Leiden, The Netherlands}
\author{H. T. Quan (全海涛)\orcidlink{0000-0002-4130-2924}}
\email{htquan@pku.edu.cn}
\affiliation{School of Physics, Peking University, Beijing, 100871, China}
\affiliation{Collaborative Innovation Center of Quantum Matter, Beijing 100871, China}
\affiliation{Frontiers Science Center for Nano-optoelectronics, Peking University,
Beijing, 100871, China}

\maketitle
\end{CJK*}

\section{the uniqueness of quasiprobability distribution}
We mention in the main text that our definition of the mixed Wigner and Margenau-Hill (WMH) distribution can be uniquely determined by the following properties: 
(1) Under Lorentz transformation $x^\prime{}^\mu =\Lambda^\mu_\nu x^\nu$, the distribution functions in two frames satisfy $P_\text{WMH}(\{w^\mu\})=P^\prime_\text{WMH}(\{\Lambda^\mu_\nu w^\nu\})$. 
(2) The marginal distribution of the energy component $W^0$ reduces to the Margenau-Hill (MH) quasiprobability distribution of work.
The first property is a requirement of the principle of covariance 
while the second property is according to the previous study on the quasiprobability of quantum work \cite{Pei2023}. 

Let us write down the above requirements in terms of the characteristic function.
The first requirement can be written as 
\begin{equation}
    \chi(\{\tilde w_\mu\}) = \chi^\prime(\{(\Lambda^{-1})^\nu_\mu \tilde w_\nu\}),
\end{equation}
in which $\Lambda^{-1}$ is the inverse of the Lorentz transformation matrix.
The second requirement implies that 
the marginal characteristic function for $W^0$ reduces to the MH one, 
\begin{equation}
    \chi(\tilde w_0,0,0,0) = \frac12 \Tr [\exp({\ui \tilde w_0 H_+})\exp({-\ui \tilde w_0 H_-})\rho] +\text{s.o.}.
\end{equation}
Here, $\text{s.o.}$ is a shorthand notation for the symmetrized ordering of the previous term, i.e., $\Tr[AB\rho]+\text{s.o.}=\Tr[AB\rho]+\Tr[BA\rho]$ for generic operators $A$ and $B$. 
Note that the above relation for the marginal characteristic function holds in any reference frame. 

We consider the characteristic function in the lab frame $\chi(\{s_\mu\})$. 
For any timelike $s_\mu$ satisfying $s_\mu s^\mu >0$, where we denote by $s=\sqrt{s_\mu s^\mu}>0$. 
There exists a Lorentz transformation $\Lambda$, such that only the 0th component is nonzero in the new frame, $s^\prime_\mu =(\Lambda^{-1})^\nu_\mu s_\nu= (s,0,0,0)$. 
Then, the characteristic functions in two frames are related by $\chi(\{s_\mu\})=\chi^\prime(s,0,0,0)$.
The characteristic function in the new frame reduces to the MH form,
\begin{equation}
    \chi^\prime (s,0,0,0) = \frac12 \Tr [\exp({\ui s H_+^\prime})\exp({-\ui s H_-^\prime})\rho] +\text{s.o.},
\end{equation}
where $H_\mp^\prime$ are the initial and final Hamiltonians in the new frame. 
We notice the relation
\begin{equation}
    s H^\prime = s_\mu P^\mu ,
\end{equation}
and then rewrite the previous formula,
\begin{equation}\label{c_WMH}
    \chi(\{s_\mu\})  = \frac12 \Tr [\exp({\ui s_\mu P_+^\mu})\exp({-\ui s_\mu P^\mu_-})\rho] +\text{s.o.}.
\end{equation}
Therefore, the characteristic function for a timelike counting field is uniquely defined.

For spacelike $s^\mu$ with $s^\mu s_\mu< 0$, a similar derivation can be made if we assume that the marginal distribution of the momentum component of the work 4-vector is also characterized by the MH distribution. 
For  lightlike $s^\mu$ with $s^\mu s_\mu=0 $, because of the continuity of the characteristic function, 
we know that Eq.~\eqref{c_WMH} also holds.

\section{proof of covariant fluctuation theorem}
In the main text, we mention that the covariant Crooks relation is generally not satisfied at the level of the joint distribution,
but it is satisfied at the level of the distribution of a specific linear combination $u_\mu W^\mu$. 
That is, 
\begin{align}
    &\frac{P_\text{WMH}(\{w^\mu\})}{\bar P_\text{WMH}(\{-w^\mu\})} \neq \exp\left( \beta_\mu w^\mu -\beta\Delta F\right),\\
    &\frac{P_{u_\mu W^\mu}(w)}{\bar P_{u_\mu W^\mu}(-w)} = \exp\left( \beta w -\beta\Delta F\right).
\end{align}
In terms of the characteristic function, the above two equations are written equivalently as 
\begin{align}
    &\chi_\text{WMH}(\{\tilde w_\mu\}) \neq \bar\chi_\text{WMH}(\{-\tilde w_\mu +\ui \beta_\mu\}) \exp(-\beta \Delta F),
\\
    &\chi_\text{WMH}(\{s u_\mu\}) = \bar\chi_\text{WMH}(\{- s u_\mu +\ui \beta_\mu\}) \exp(-\beta \Delta F).
\end{align}
In the following, we derive these two formulas. 

We recall our setup: 
an isolated quantum field $\phi$ is driven by an external driving field $h$. 
The driving process starts from the initial hypersurface $\Sigma_-$ and ends on the final hypersurface $\Sigma_+$.
The energy-momentum tensor $T^{\mu\nu}(\phi,\partial \phi,h)$ is a function of the field and its derivatives. 
The 4-momentum on the initial and final hypersurfaces are given by 
\begin{equation}
    P^\nu_\mp = \int_{\Sigma_\mp} \ud \sigma n_\mu T^{\mu\nu}(\phi,\partial \phi,h).
\end{equation}
The initial state is a moving equilibrium state $\rho^\text{eq} = \exp(-\beta_\mu P^\mu_-)/Z_-$.

The action of the field is denoted by $I[\phi,h]$.
The time evolution satisfies the following Euler-Lagrange equation for operators, 
\begin{equation}
    \delta I[\phi,h] =0.
\end{equation}
Note that the above field equation should be understood as an equation for operators in the Heisenberg picture.

We suppose that the system satisfies the PT symmetry, 
which implies $I[\phi(-x),h(-x)]=I[\phi(x),h(x)]$. 
In the backward process, the external field is reversed $\bar h(x)=h(-x)$.
Then, $\bar\phi(x)=\phi(-x)$ is a solution of the Euler-Lagrange operator field equation in the backward process. 
Thus, we may take $\bar\phi(x)$ as the time-evolution of the field operator in the backward process. 

We note that the energy-momentum tensor does not change for the spacetime-reversal field, $\bar T^{\mu\nu}(-x) = T^{\mu\nu}(x)$.
Therefore, the 4-momentum in the backward process satisfies the following relation 
\begin{equation}\label{pm}
    \bar P_-^\mu = P_+^\mu , \quad\bar P_+^\mu = P_-^\mu .
\end{equation}
This implies the initial and final 4-momentum are exchanged in the backward process.

The characteristic function for the WMH quasiprobability of the work 4-vector in the forward process is 
\begin{equation}
    \chi_\text{WMH}(\{\tilde w_\mu\})=\frac12\Tr \left[\exp({\ui \tilde w_\mu P^\mu_+})\exp({-\ui \tilde w_\mu P^\mu_-})\exp({-\beta_\mu P^\mu_-})/Z_- \right]+\text{s.o.} ,
\end{equation}
while the characteristic function in the backward process is 
\begin{equation}
    \bar\chi_\text{WMH}(\{\tilde w_\mu\})=\frac12\Tr \left[\exp({\ui \tilde w_\mu\bar P^\mu_+})\exp({-\ui \tilde w_\mu\bar P^\mu_-})\exp({-\beta_\mu\bar P^\mu_-})/Z_+ \right] +\text{s.o.}.
\end{equation}
We investigate the following quantity and utilize the relation in Eq.~\eqref{pm}
\begin{equation}\label{crookschi}
   \begin{aligned}
\bar\chi_\text{WMH}(\{-\tilde w_\mu+\ui \beta_\mu\})= \frac12\Tr \left[\exp({-\beta_\mu P^\mu_--\ui \tilde w_\mu  P^\mu_-})\exp({\ui \tilde w_\mu P^\mu_++\beta_\mu P^\mu_+})\exp({-\beta_\mu P^\mu_+}) /Z_+\right]\\+\text{s.o.}
       \end{aligned}
\end{equation}
Due to the noncommutativity of energy and momentum, the summation of operators in the exponential cannot be split for general $\tilde w_\mu$, 
\begin{equation}
    \exp(\ui \tilde w_\mu P^\mu_++\beta_\mu P^\mu_+) \neq \exp(\ui\tilde w_\mu P^\mu_+)\exp(\beta_\mu P^\mu_+).
\end{equation}
Therefore, we cannot relate Eq.~\eqref{crookschi} to $\chi(\{\tilde w_\mu\})$, and the Crooks relation is not satisfied at the level of joint distribution,
\begin{equation}
    \bar\chi_\text{WMH}(\{-\tilde w_\mu+\ui \beta_\mu\})\exp(-\beta\Delta F)\neq\chi_\mathrm{WMH}(\{\tilde w_\mu\}).
\end{equation}
On the other hand, when we choose $\tilde w_\mu =s u_\mu $ which is parallel to $\beta_\mu$, the operators $ su_\mu P^\mu_\mp$ and $\beta_\mu P^\mu_\mp$ commute.
In this case, we can simplify Eq.~\eqref{crookschi} into
\begin{equation}
       \begin{aligned}
&\bar\chi_\text{WMH}(\{-s u_\mu+\ui \beta_\mu\})\\&= \frac12\Tr \left[\exp(-\beta_\mu P^\mu_--\ui  su_\mu  P^\mu_-)\exp(\ui su_\mu P^\mu_++\beta_\mu P^\mu_+)\exp(-\beta_\mu P^\mu_+) /Z_+\right]+\text{s.o.}\\
&=\frac12 \Tr\left[\exp(-\beta_\mu P^\mu_-)\exp(-\ui  su_\mu  P^\mu_-)\exp(\ui su_\mu P^\mu_+) /Z_+\right]+\text{s.o.}\\
&=\chi_\text{WMH}(\{s u_\mu\}) \exp(\beta\Delta F).
       \end{aligned}
\end{equation}
Here, we have used $\exp(-\beta\Delta F)=Z_+/Z_-$.

We see clearly that due to the noncommutativity, the covariant quantum Crooks relation is violated at the level of joint distribution, but it is valid for the distribution of a specific linear combination of energy and momentum components of work 4-vector, $u_\mu W^\mu$.  
At last, the Crooks relation for the distribution of this specific linear combination implies the covariant Jarzynski equality, 
\begin{equation}
    \ev{\exp(-\beta_\mu W^\mu )}=\exp(-\beta\Delta F),
\end{equation}
since the Jarzynski equality only involves the linear combination $u_\mu W^\mu$. 

\section{Proof of covariant Crooks fluctuation theorem for open quantum system}
For an open quantum system (field $\phi$) coupled to a heat bath (field $\psi$) and driven by an external field $h$, the system and the bath together form a closed system. 
The field operators of the system and the bath commute with each other, $[\psi,\phi]=0$. 
The total action of the composite system is
\begin{equation}
    I[\phi,\psi,h]=I_\text{s}[\psi,h]+I_\text{b}[\psi]+\varepsilon I_{int}[\psi,\phi],
\end{equation}
Here $\varepsilon I_{int}$ indicates that the coupling between the field and the system is weak; $I_\text{s}$ and $I_\text{b}$ denote the action contributed by the system and the bath.
In the Heisenberg picture, the time evolution of the composite system satisfies the Euler-Lagrange equation for field operators,
\begin{equation}
    \delta I[\phi,\psi,h]=0.
\end{equation}
Similar to the previous section, we assume the PT symmetry, so that $I[\phi(-x),\psi(-x),h(-x)]=I[\phi(x),\psi(x),h(x)]$. In the backward process, the external field \replaced{undergoes a spacetime reversal}{is reversed} $\bar h(x)=h(-x)$\replaced{. Then,}{, then,} $\bar\phi(x)=\phi(-x),~\bar \psi(x)=\psi(-x)$ is also a solution of the Euler-Lagrange equation \replaced{in}{is} the backward process.

From the actions $I_\text{s}$ and $I_\text{b}$, we can derive the energy-momentum tensor $T^{\mu\nu}_{s}(\phi,\partial\phi,h)$ of the system and $T^{\mu\nu}_\text{b} (\psi,\partial \psi)$ of the bath (the contribution from the interaction term is negligible).
The energy-momentum tensors do not change for the spacetime-reversal process, $\bar T^{\mu\nu}_\text{s}(-x)=T^{\mu\nu}_{s}(x)$, $\bar T^{\mu\nu}_\text{b} (-x)=T^{\mu\nu}_\text{b} (x)$.
Therefore, the 4-momentum in the backward process satisfies the following relation:
\begin{equation}\label{pm open}
    \bar P_{\text{s}-}^\mu = P_{\text{s}+}^\mu , ~\bar P_{\text{s}+}^\mu = P_{\text{s}-}^\mu,\quad \bar P_{\text{b}-}^\mu = P_{\text{b}+}^\mu , ~\bar P_{\text{b}+}^\mu = P_{\text{b}-}^\mu.
\end{equation}
Similar relations also hold for the total 4-momentum of the composite system $P^\mu_{t}=P^\mu_{s}+P^\mu_{b}$.
In the weak coupling regime, the 4-momentum contributed by the interaction term is negligible.
Due to the commutation relation of the field operator, the four-momentum operators satisfy the commutation relation $[P^\mu_\text{s}, P^\nu_\text{b}]=0$.

In our setup, we assume that  in the forward process, 
the system and bath are initially moving equilibrium states with different inverse temperature 4-vectors $\beta^\mathrm{s}_\mu$ and $\beta^\mathrm{b}_\mu$. 
The total density operator $\rho$ is the following product state, 
\begin{equation}
    \rho=\exp(-\beta^\text{s}_\mu P^\mu_{\text{s}-})/Z_{\text{s}-} \otimes \exp(-\beta^\text{b}_\mu P^\mu_{\text{b}-})/Z_{b}.
\end{equation}
In the backward process, 
the system and bath have the same inverse temperature 4-vectors as in the forward process, 
and the total density operator is
\begin{equation}
    \bar \rho=\exp(-\beta^\text{s}_\mu P^\mu_{\text{s}+})/Z_{\text{s}+} \otimes \exp(-\beta^\text{b}_\mu P^\mu_{\text{b}+})/Z_{b}.
\end{equation}
The difference between two partition functions $Z_\mathrm{s-}$ and $Z_\mathrm{s+}$ is because of the different initial configurations of the external driving field $h$.
The joint characteristic function for the WMH quasiprobability of the work and heat 4-vectors in the
forward process is
\begin{equation}
    \chi_\mathrm{WMH}(\{\tilde w_\mu\},\{\tilde q_\mu\}) = \frac12\Tr[\exp({\ui \tilde w_\mu P^\mu_{\text{t}+}-\ui \tilde q_\mu P^\mu_{\text{b}+}})\exp({-\ui \tilde w_\mu P^\mu_{\text{t}-}+\ui \tilde q_\mu P^\mu_{\text{b}-}})\rho]+\text{s.o.}.
\end{equation}
For the backward process, we have
\begin{equation}
   \bar \chi_\mathrm{WMH}(\{\tilde w_\mu\},\{\tilde q_\mu\}) = \frac12\Tr[\exp({\ui \tilde w_\mu \bar P^\mu_{\text{t}+}-\ui \tilde q_\mu \bar P^\mu_{\text{b}+}})\exp({-\ui \tilde w_\mu \bar P^\mu_{\text{t}-}+\ui \tilde q_\mu \bar P^\mu_{\text{b}-}{\text{b}-}})\bar\rho]+\text{s.o.}.
\end{equation}
In the above formulas, $\text{s.o.}$ also represents the term from symmetric ordering.

\replaced{As is mentioned in the main text}{Due to the non-commutativity}, the covariant Crooks fluctuation theorem for open quantum systems is only valid for two specific linear combinations of $W^\mu$ and $Q^\mu$: $u^\text{s}_\mu(W^\mu+Q^\mu)$ and $u^\text{b}_\mu Q^\mu$,
\begin{equation}\label{crooks open}
\begin{aligned}
    \frac{P_{u_\mu^\text{s} (W^\mu +Q^\mu),u_\mu^\text{b}Q^\mu}(u_\mu^\text{s}(w^\mu+q^\mu),u_\mu^\text{b}q^\mu)}{\bar P_{u_\mu^\text{s} (W^\mu +Q^\mu),u_\mu^\text{b}Q^\mu}(-u_\mu^\text{s}(w^\mu+q^\mu),-u_\mu^\text{b}q^\mu)} 
    = \exp\left[\beta_\mu^\text{s} w^\mu-\beta^\mathrm{s}\Delta F-(\beta^\text{b}_\mu-\beta^\text{s}_\mu)q^\mu\right].
\end{aligned}
\end{equation}
In terms of the characteristic function, the above relation reads as follows,
\begin{equation}\label{crooks open chara}
    \begin{aligned}
        &\chi_\mathrm{WMH}(\{\lambda_1 u^\text{s}_\mu\},\{\lambda_1u^\text{s}_\mu-\lambda_2u^\text{b}_\mu\})\\
        =&\bar \chi_\mathrm{WMH}(\{(-\lambda_1 +\ui\beta^\text{s})u^\text{s}_\mu\},\{(-\lambda_1+\ui\beta^\text{s})u^\text{s}_\mu+(\lambda_2-\ui\beta^\text{b})u^\text{b}_\mu\})\exp(-\beta^\text{s}\Delta F).
    \end{aligned}
\end{equation}
The proof of Eq.~\eqref{crooks open chara} is as follows:
\begin{equation}
\begin{aligned}
    &\bar \chi_\mathrm{WMH}(\{(-\lambda_1 +\ui\beta^\text{s})u^\text{s}_\mu\},\{(-\lambda_1+\ui\beta^\text{s})u^\text{s}_\mu+(\lambda_2-\ui\beta^\text{b})u^\text{b}_\mu\})\\
    =& \frac12\Tr[\exp({\ui (-\lambda_1 +\ui\beta^\text{s})u^\text{s}_\mu \bar P^\mu_{\text{t}+}-\ui (-\lambda_1 +\ui\beta^\text{s})u^\text{s}_\mu \bar P^\mu_{\text{b}+}-\ui (\lambda_2-\ui\beta^\text{b})u^\text{b}_\mu \bar P^\mu_{\text{b}+}})\\
    &\quad \quad \exp({-\ui (-\lambda_1 +\ui\beta^\text{s})u^\text{s}_\mu \bar P^\mu_{\text{t}-}
    +\ui (-\lambda_1 +\ui\beta^\text{s})u^\text{s}_\mu \bar P^\mu_{\text{b}-}
    +\ui (\lambda_2-\ui\beta^\text{b})u^\text{b}_\mu \bar P^\mu_{\text{b}-}})\bar\rho]+\text{s.o.}\\
    =& \frac12\Tr[\exp({\ui (-\lambda_1 +\ui\beta^\text{s})u^\text{s}_\mu  P^\mu_{\text{s}-}-\ui (\lambda_2-\ui\beta^\text{b})u^\text{b}_\mu  P^\mu_{\text{b}-}})\\
    &\quad\quad\exp({-\ui (-\lambda_1 +\ui\beta^\text{s})u^\text{s}_\mu P^\mu_{\text{s}+}
    +\ui (\lambda_2-\ui\beta^\text{b})u^\text{b}_\mu P^\mu_{\text{b}+}})\bar\rho]+\text{s.o.}\\
    =& \frac12\Tr[\exp(-\beta^\text{s}_\mu P^\mu_{\text{s}-}-\beta^\text{b}_\mu P^\mu_{\text{b}-})\exp({-\ui\lambda_1u^\text{s}_\mu  P^\mu_{\text{s}-}-\ui \lambda_2 u^\text{b}_\mu  P^\mu_{\text{b}-}})\\
    &\quad\quad\exp({\ui \lambda_1 u^\text{s}_\mu P^\mu_{\text{s}+}
    +\ui \lambda_2 u^\text{b}_\mu P^\mu_{\text{b}+}})\exp(\beta^\text{s}_\mu P^\mu_{\text{s}+}+\beta^\text{b}_\mu P^\mu_{\text{b}+})\bar\rho]+\text{s.o.}\\
    =& \frac12 \frac{Z_{\text{s}-}Z_\text{b}}{Z_{\text{s}+}Z_\text{b}}\Tr[\rho \exp({-\ui\lambda_1u^\text{s}_\mu  P^\mu_{\text{s}-}-\ui \lambda_2 u^\text{b}_\mu  P^\mu_{\text{b}-}})\exp({\ui \lambda_1 u^\text{s}_\mu P^\mu_{\text{s}+}
    +\ui \lambda_2 u^\text{b}_\mu P^\mu_{\text{b}+}}) ]+\text{s.o.}\\
    =& \frac12\Tr[\rho \exp({-\ui\lambda_1u^\text{s}_\mu  P^\mu_{\text{t}-}+\ui (\lambda_1u^\text{s}_\mu-\lambda_2 u^\text{b}_\mu )  P^\mu_{\text{b}-}})\\
    &\quad\quad\exp({\ui \lambda_1 u^\text{s}_\mu P^\mu_{\text{t}+}
    -\ui (\lambda_1u^\text{s}_\mu-\lambda_2 u^\text{b}_\mu ) P^\mu_{\text{b}+}})]\exp(\beta^\text{s}\Delta F)+\text{s.o.}\\
    =&\chi_\mathrm{WMH}(\{\lambda_1 u^\text{s}_\mu\},\{\lambda_1u^\text{s}_\mu-\lambda_2u^\text{b}_\mu\}) \exp(\beta^{\mathrm{s}}\Delta F).
\end{aligned}
\end{equation}The first equality is the definition.
In the second equality, we use the relation of the operator in the forward and backward processes Eq.~\eqref{pm open}.
In the third equality, we use the fact that $P_{
\mathrm{s\pm}}^{\mu}
$ and $P_{
\mathrm{b\pm}}^{\mu} 
$ commute. We expand the result and rewrite it with $\rho$ and the partition function in the fourth equality.  
In the fifth equality, we use $P^\mu_{\mathrm{s}\pm}= P_{\mathrm{t}\pm}^\mu -P_{\mathrm{b}\pm}^\mu$ and $\exp(\beta^\mathrm{s}\Delta F)=Z_{\mathrm{s}-}/Z_{\mathrm{s}+}$.

From Eq.~\eqref{crooks open}, we can see that the fluctuation theorem is for the joint distribution of the change of $u^\text{s}_\mu  P^\mu_\text{s}$ and $u^\text{b}_\mu  P^\mu_\text{b}$, which respectively correspond to the energy of the system measured in the rest frame of the system and the energy of the bath measured in the rest frame of the bath.
This joint distribution is positive definite and reduces to the result of a joint two-projective-measurement scheme of both operators $u^\text{s}_\mu  P^\mu_\text{s}$ and $u^\text{b}_\mu  P^\mu_\text{b}$.

\section{Example}
In this section, we investigate the example of a quantum scalar field driven by a beam of classical particles, as described in the main text.
Here, we consider the problem in ($d+1$)D.
The 4-momentum operator is $P^\mu=(P^0, P^i)$.
The Lagrangian of the system is 
\begin{equation}
    \mathcal L = \frac1 2 \partial^\mu \phi \partial_\mu\phi - \frac12 m^2 \phi^2 -h\phi.
\end{equation}
In the rest reference frame of the initial thermal state, the particles are initially at rest, with a Gaussian distribution of the particle number density,
\begin{align}
n(\bm{l})=\frac{N_{0}}{(\sqrt{2\pi}\sigma)^{d}}\exp(-\bm{l}^{2}/2\sigma^{2}).
\end{align}
Here $N_0$ is the total number of particles, and $\sigma$ is the spread of the beam.
In the reference frame of the initial thermal state, when the particles intersect the initial hypersurface $\mathrm{\Sigma}_{-}$ (the isochronous surface $t=0$), each particle in the beam instantaneously acquires a velocity $\bm v_b$.
Subsequently, when the particles meet the final hypersurface $\mathrm{\Sigma}_{+}$ (the isochronous surface $t=T$), they become static again in the same frame.
The beam generates a classical external field
\begin{equation}
    h(x^\mu)=g\int \ud\bm l \int \ud\tau~n(\bm l) \delta^{d+1}[x^\mu-x^\mu_{\bm l}(\tau)].
\end{equation}
For a classical particle with initial position $\bm l$, the worldline of the particle is
\begin{equation}
        x^\mu_{\bm l}(\tau)
    =\begin{cases}
        (\tau,\bm l) ,&\tau<0;\\
        (\frac{\tau}{\sqrt{1-v_b^2}},\bm l+\bm v_b \frac{\tau}{\sqrt{1-v_b^2}}),& 0\leq\tau\leq T\sqrt{1-v_b^2};\\
        (\tau+T-T\sqrt{1-v_b^2},\bm l+\bm v_b T) ,& \tau >T\sqrt{1-v_b^2}.
    \end{cases}
\end{equation}
For $d=1$, the worldline expressed in terms of the proper time exactly corresponds to the worldline expressed with the time coordinate $t$ in Eq.~(14) of the main text.
Here, we use $\tau$ instead of $t$ because from the above formula, $h$ could be directly calculated.

The equation of motion can be derived from the Lagrangian as
\begin{equation}
    \partial^\mu\partial_\mu\phi+m^2\phi=-h(x).
\end{equation}
The solution of the equation of motion is
\begin{equation}\label{eq:solution of KG eqn}
    \phi(x)=\phi_0(x)-\int_0^{x^0}\ud y^0\int \ud^dy^i G_\mathrm{R}(x-y)h(y),
\end{equation}
where $\phi_0$ is the solution of the free field,
\begin{equation}
    \phi_0(x) = \int\frac{\ud^d k}{(2\pi)^d 2\omega_k} \left[ \ue^{-\ui k^\mu x_\mu}a(\bm k ) + \ue^{\ui k^\mu x_\mu}a^\dagger(\bm k ) \right],
\end{equation}
with creation and annihilation operators $a^\dagger$ and $a$, satisfying $\comm{a(\bm k)}{a^\dagger(\bm k^\prime)}=(2\pi)^d 2\omega_{ k}\delta^d(\bm k-\bm k^\prime) $.
$G_\mathrm{R}(x-y)$ is the retarded Green's function of the classical Klein-Gordon field,
\begin{equation}
    G_\mathrm{R}(x) = \ui\theta(x^0)\int\frac{\ud^d k}{(2\pi)^d 2\omega_k} \left( \ue^{-\ui k^\mu x_\mu}-\ue^{\ui k^\mu x_\mu} \right),
\end{equation}
where $\theta(x^0)$ is the Heaviside function.

The energy-momentum tensor can be obtained from the Lagrangian,
\begin{equation}
     T^{\mu\nu} = \frac{\partial\mathcal L}{\partial\partial_\mu\phi}\partial^\nu\phi- \eta^{\mu\nu} \mathcal L.
\end{equation}
With the solution of the nonhomogeneous equation, we can calculate the 4-momentum operator.
On the initial hypersurface $\mathrm{\Sigma}_{-}$, the initial value of the field is chosen as the operator of the free field, $\phi(0,x)=\phi_0(0,x)$. 
Therefore, the initial 4-momentum is 
\begin{equation}\label{eq:solution of initial four momentum}
    P^0_- = P^0_0 + \int\ud^d x ~h(0,\bm x)\phi(0,\bm x), \quad P^i_- = P^i_0,
\end{equation}
where $P^\mu_0$ are the 4-momentum of the free field, 
\begin{equation}
 P^\mu_0= \int\frac{\ud^d k}{(2\pi)^d 2\omega_k} k^\mu a^\dagger(\bm k)a(\bm k).
\end{equation}
By using the Fourier space, 
\begin{equation}
    h(0,\bm x) = \int\frac{\ud^d k}{(2\pi)^d}\ue^{\ui \bm k\cdot\bm x}h_0(\bm k),
\end{equation}
the initial Hamiltonian is 
\begin{equation}\label{eq:solution of initial energy}
    P^0_- = P^0_0 + \int\frac{\ud ^dk}{(2\pi)^d 2\omega_k} \left[a(\bm k) h_0(\bm k)^* +a^\dagger(\bm k) h_0(\bm k) \right].
\end{equation}
The final field operator is 
\begin{equation}
\begin{aligned}
    \phi(\tau,\bm x) &= \phi_0(x) - \int_0^{\tau}\ud y^0\int\ud^d \bm y~\ui\theta(\tau-y^0)\int\frac{\ud^d k}{(2\pi)^d 2\omega_k} \left( \ue^{-\ui k^\mu (x_\mu-y_\mu)}-\ue^{\ui k^\mu (x_\mu-y_\mu)} \right) h(y)  \\
    &=\phi_0(x)-\ui \int\frac{\ud^d k}{(2\pi)^d 2\omega_k} \ue^{\ui \bm k\cdot \bm x} \left[\ue^{-\ui\omega_k \tau}h(\bm k,\omega_k)-\ue^{\ui\omega_k\tau}h(\bm k,-\omega_k)\right],
    \end{aligned}
\end{equation}
where 
\begin{equation}
    h(\bm k,\omega) = \int_0^\tau\ud t\int\ud^d x ~\ue^{\ui\omega t-\ui \bm k\cdot\bm x} h(t,\bm x).
\end{equation}
The 4-momentum on the final hypersurface $\mathrm{\Sigma}_{+}$ can then be expressed as
\begin{equation}\label{eq:solution of final four momentum}
    \begin{split}
        P^0_+=&P^0_0+ \int\frac{\ud^d k}{(2\pi)^d 2\omega_k} \{ \omega_k~[\ui h(\bm k,\omega_k)^* a(\bm k)+h.c.]
        + \omega_k~[h(\bm k,\omega_k)^* h(\bm k,\omega_k)]\\
        &+[\ue^{-\ui\omega_k T} h_T(\bm k)^*a(\bm k)+h.c.] 
        +[\ui \ue^{-\ui\omega_k T} h_T(\bm k)^*h(\bm k,\omega_k)+h.c.] \},\\
        P^i_+=&P^i_0+\int\frac{\ud^d k}{(2\pi)^d 2\omega_k} \{ k^i~[\ui h(\bm k,\omega_k)^* a(\bm k)+h.c.]
        +k^i~[h(\bm k,\omega_k)^* h(\bm k,\omega_k)]\}.
    \end{split}
\end{equation}
Here, $h_t(\bm k)=\int dx~\ue^{-i\bm k \cdot \bm x}h(t,\bm x)$ is the spatial Fourier transformation of $h$ on a spacelike hypersurface with time coordinate $t$.

The 4-momentums of initial and final state are thus expressed as a shifted quadratic form that may contain linear or constant terms
\begin{equation}
    P^\mu_{\pm}(\bm k)=k^\mu a^\dagger(\bm k)a(\bm k)+A^\mu_{\pm}a(\bm k)+A^{*\mu}_{\pm}a^\dagger(\bm k)+B^\mu_{\pm}(\bm k).
\end{equation}
The contribution of a single mode $\bm k$ to the WMH characteristic function is
\begin{equation}\label{trace_tobec}
    \chi_k(\{\tilde{w}_\mu\})=\frac 12\Tr\left[\ue^{\ui \tilde w_\mu P_+^\mu(\bm k)} \ue^{-\ui \tilde w_\mu P_-^\mu(\bm k)}\ue^{-\beta_\mu P^\mu_-(\bm k)}/Z_{-,\bm k}\right]+ \text{s.o.}.
\end{equation}
Different modes independently contribute to the total work, and the logarithm of the total characteristic function can be calculated from
\begin{equation}
    \log\chi_\text{WMH}(\{\tilde{w}_\mu\})=\int \frac{\ud^dk}{(2\pi)^d 2\omega_k} \log(\chi_k(\{\tilde{w}_\mu\})).
\end{equation}
For a single mode, we can calculate the characteristic function of work using the trace formula given in the End Matter. 
In our model, the Nambu basis is simply $A=(a(\bm k),a^\dagger(\bm k))$.
Using the quadratic forms given in Eqs.~\eqref{eq:solution of initial four momentum}, \eqref{eq:solution of initial energy}, and \eqref{eq:solution of final four momentum}, we obtain the explicit result of Eq.~\eqref{trace_tobec} as 
{\footnotesize
\begin{equation}
\begin{aligned}
    &\log \chi_k(\{\tilde{w}_\mu\})= 
    \frac12\ue^{-\ui \tilde{w}_\mu k^\mu}[\beta_\mu k^\mu (\tilde{w}_\mu k^\mu)^2 (\ue^{\beta_\mu k^\mu}-1)]^{-1} [\beta_0 \tilde{w}_\mu k^\mu (\ue^{(\beta_\mu+\ui\tilde{w}_\mu)k^\mu}\\
    &\times(\tilde{w}_0 (\ue^{\ui \omega_k T} h^*_0(\bm{k}) h_T(\bm{k})-\ue^{-\ui \omega_k T} h_0(\bm{k}) h^*_T(\bm{k}))    -\ui \tilde{w}_\mu k^\mu (h(\bm k,\omega_k) h^*_0(\bm{k})+h_0(\bm{k}) h^*(\bm k,\omega_k)))\\
    &+h^*_0(\bm{k}) \ue^{(\beta_\mu+2\ui\tilde{w}_\mu)k^\mu} (\tilde{w}_0 (h_0(\bm{k})-\ue^{\ui \omega_k T} h_T(\bm{k}))+\ui \tilde{w}_\mu k^\mu h(\bm k,\omega_k))\\
   &+h_0(\bm{k}) \ue^{\beta_\mu k^\mu} (\ue^{-\ui \omega_k T} \tilde{w}_0 h^*_T(\bm{k})+\ui \tilde{w}_\mu k^\mu h^*(\bm k,\omega_k)-\tilde{w}_0 h^*_0(\bm{k}))\\
   &+\ue^{\ui \tilde{w}_\mu k^\mu} (-\ue^{\ui \omega_k T} \tilde{w}_0 h^*_0(\bm{k})
   h_T(\bm{k})+\ue^{-\ui \omega_k T} \tilde{w}_0 h_0(\bm{k}) h^*_T(\bm{k})+\ui \tilde{w}_\mu k^\mu (h(\bm k,\omega_k) h^*_0(\bm{k})+h_0(\bm{k}) h^*(\bm k,\omega_k)))\\
   &+h^*_0(\bm{k}) \ue^{2 \ui \tilde{w}_\mu k^\mu} (\ue^{\ui \omega_k T} \tilde{w}_0 h_T(\bm{k})-\ui \tilde{w}_\mu k^\mu h(\bm k,\omega_k)-\tilde{w}_0 h_0(\bm{k}))\\
   &+h_0(\bm{k})
   (\tilde{w}_0 (h^*_0(\bm{k})-\ue^{-\ui \omega_k T} h^*_T(\bm{k}))-\ui \tilde{w}_\mu k^\mu h^*(\bm k,\omega_k)))\\
   &+\beta_\mu k^\mu (\tilde{w}_0^2 (\ue^{\ui \tilde{w}_\mu k^\mu}-1) (-\ue^{-\ui \omega_k T} h^*_T(\bm{k}) \ue^{(\beta_\mu+\ui\tilde{w}_\mu)k^\mu} (h_0(\bm{k})-\ue^{\ui \omega_k T}
   h_T(\bm{k}))\\
   &-h_0(\bm{k}) \ue^{\beta_\mu k^\mu} (h^*_0(\bm{k})-\ue^{-\ui \omega_k T} h^*_T(\bm{k}))\\
   &+\ue^{\ui \omega_k T} h_T(\bm{k}) (h^*_0(\bm{k})-\ue^{-\ui \omega_k T} h^*_T(\bm{k}))+h^*_0(\bm{k}) \ue^{\ui \tilde{w}_\mu k^\mu} (h_0(\bm{k})-\ue^{\ui \omega_k T} h_T(\bm{k})))\\
   &-\ui \tilde{w}_\mu k^\mu \tilde{w}_0
   (\ue^{(\beta_\mu+\ui\tilde{w}_\mu)k^\mu} (\tilde{w}_0 h_T(\bm{k}) h^*_T(\bm{k})+\ue^{\ui \omega_k T} h_T(\bm{k}) h^*(\bm k,\omega_k)-\tilde{w}_0 h_0(\bm{k}) h^*_0(\bm{k})-2 h_0(\bm{k}) h^*(\bm k,\omega_k))\\
   &+h^*(\bm k,\omega_k) \ue^{(\beta_\mu+2\ui\tilde{w}_\mu)k^\mu} (h_0(\bm{k})-\ue^{\ui \omega_k T}h_T(\bm{k}))\\
   &-(h(\bm k,\omega_k) (\ue^{\ui \tilde{w}_\mu k^\mu}-1) (h^*_0(\bm{k}) (\ue^{\ui \tilde{w}_\mu k^\mu}-1)-\ue^{-\ui \omega_k T} h^*_T(\bm{k}) (\ue^{(\beta_\mu+\ui\tilde{w}_\mu)k^\mu}-1)))+h_0(\bm{k}) h^*(\bm k,\omega_k) \ue^{\beta_\mu k^\mu}\\
   &+\ue^{\ui \tilde{w}_\mu k^\mu} (-\ue^{\ui \omega_k T} \ue^{-\ui \omega_k T} \tilde{w}_0 h_T(\bm{k}) h^*_T(\bm{k})+\ue^{\ui \omega_k T} h_T(\bm{k}) h^*(\bm k,\omega_k)+\tilde{w}_0 h_0(\bm{k}) h^*_0(\bm{k}))-\ue^{\ui \omega_k T} h_T(\bm{k}) h^*(\bm k,\omega_k))\\
   &+(\tilde{w}_\mu k^\mu)^2 h(\bm k,\omega_k) h^*(\bm k,\omega_k) (\ue^{\ui \tilde{w}_\mu k^\mu}-1) (\ue^{(\beta_\mu+\ui\tilde{w}_\mu)k^\mu}-1))]\\
   &+\text{s.o.},
   \end{aligned}
\end{equation}
}which is used to evaluate the characteristic function of work in Fig.~1 of the main text.

\bibliography{cqft}